\documentclass[journal,twocolumn]{IEEEtran}

\usepackage{amsmath}
\usepackage{bbm}
\usepackage{epsfig}
\usepackage{algpseudocode}
\usepackage{stfloats}
\usepackage{algorithm}

\newcommand{\myvec}[1]{ \mathbf{#1} }
\newcommand{\myvecsymb}[1]{ \boldsymbol{#1} }
\newcommand{\transp}[1]{ {#1}^T }
\newcommand{\mymatrix}[1]{ \mathbf{#1} }
\DeclareMathOperator*{\argmax}{arg\,max}
\newcommand{\eqdef}{\stackrel{\triangle}{=}}
\newcommand{\myfigurewidth}{.99\columnwidth}

\begin{document}

\title{The Sparse Principal Component\\of a Constant-rank Matrix}%
\author{Megasthenis~Asteris,~\IEEEmembership{Student~Member,~IEEE,}
Dimitris~S.~Papailiopoulos,~\IEEEmembership{Student~Member,~IEEE,}
and~George~N.~Karystinos,~\IEEEmembership{Member,~IEEE}%
\thanks{Manuscript received May 29, 2013; accepted November 28, 2013.
The associate editor coordinating the review of this paper and approving it for publication was Dr. Yi Ma.}%
\thanks{This paper was presented at the 2011 IEEE International Symposium on Information Theory (ISIT 2011), Saint Petersburg, Russia, Aug. 2011.
This work was supported by the European Union (European Social Fund - ESF) and Greek national funds through the Operational Program ``Education and Lifelong Learning'' of the National Strategic Reference Framework (NSRF) Research Funding Program ``Thales - Investing in knowledge society through the European Social Fund.''}%
\thanks{M. Asteris and D. S. Papailiopoulos are with the Department of Electrical and Computer Engineering, The University of Texas at Austin, Austin, TX 78712, USA (e-mail: \{\texttt{megas, dimitris\}@utexas.edu}).}%
\thanks{G. N. Karystinos (corresponding author) is with the Department of Electronic and Computer Engineering, Technical University of Crete, Chania, 73100, Greece (e-mail: \texttt{karystinos@telecom.tuc.gr}).}}%

\markboth{\tiny T\MakeLowercase{his article has been accepted for publication in a future issue of} IEEE T\MakeLowercase{ransactions on} I\MakeLowercase{nformation} T\MakeLowercase{heory, but has not been fully edited.}
C\MakeLowercase{ontent may change prior to final publication.}}{}%

\IEEEpubid{\tiny Copyright~\copyright~2013 IEEE. Personal use of this material is permitted. However, permission to use this material for any other purposes must be obtained from the IEEE by sending a request to pubs-permissions@ieee.org.\hspace{2cm}}%
\maketitle

\begin{abstract}
The computation of the sparse principal component of a matrix is equivalent to the identification of its principal submatrix with the largest maximum eigenvalue.
Finding this optimal submatrix is what renders the problem ${\mathcal{NP}}$-hard.
In this work, we prove that, if the matrix is positive semidefinite and its rank is constant, then its sparse principal component is polynomially computable.
Our proof utilizes the auxiliary unit vector technique that has been recently developed to identify problems that are polynomially solvable.
Moreover, we use this technique to design an algorithm which, for any sparsity value, computes the sparse principal component with complexity ${\mathcal O}\left(N^{D+1}\right)$, where $N$ and $D$ are the matrix size and rank, respectively.
Our algorithm is fully parallelizable and memory efficient.
\end{abstract}

\begin{IEEEkeywords}
Eigenvalues and eigenfunctions,
feature extraction,
information processing,
machine learning algorithms,
principal component analysis,
signal processing algorithms.
\end{IEEEkeywords}

\section{Introduction}

Principal component analysis (PCA) is a well studied and broadly used dimensionality reduction tool.
The principal components (PCs) of a set of observations on $N$ variables capture orthogonal directions of maximum variance and offer a distance-optimal, low-dimensional representation that -for many purposes- conveys sufficient amount of information.
Without additional constraints, the PCs of a data set can be computed in polynomial time in $N$ using the eigenvalue decomposition.

A disadvantage of conventional PCA is that, in general, the extracted components are expected to have nonzero elements in all their entries.
In many applications, sparse vectors that convey information are more favorable either due to sparsity of the actual signals~\cite{Donoho2006},~\cite{Tropp2006} or because sparsity implies interpretability~\cite{zou_hastie_tibshirani_2006} when each coordinate of a PC corresponds, for example, to a different word in text analysis applications or the expression of a particular gene in bio data sets.
Thus, provided that the application requires it, some of the maximum variance of the true PCs may be traded for sparsity.
Recently, there has been an increasing interest in computing sparse components of data sets with applications that range from signal processing, communication networks, and machine learning, to bioinformatics, finance, and meteorology~\cite{moghaddam_weiss_avidan_2006b}-\cite{Wei2013}.

To enforce sparsity on the extracted components, a linearly constrained $l_0$-norm minimization problem is usually considered~\cite{Donoho2006},~\cite{Tropp2006},~\cite{Elad2009},~\cite{Tsinos2012}.
This problem is equivalent to the sparse variance maximization, that is, the maximization of the Rayleigh quotient of a matrix under an $l_0$-norm constraint on the maximizing argument~\cite{moghaddam_weiss_avidan_2006b}-\cite{Amini2008},~\cite{Shen2011},~\cite{Singh2011},~\cite{Wei2013},~\cite{moghaddam_weiss_avidan_2006a}-\cite{Grbovic2012}.
In both problems, due to the additional cardinality constraint that is enforced, the sparsity-aware flavor of PCA, termed sparse PCA, comes at a higher cost: sparse PCA is an ${\mathcal{NP}}$-hard problem~\cite{moghaddam_weiss_avidan_2006a}.

To approximate sparse PCA, various methods have been introduced in the literature.
A modified PCA technique based on the LASSO was introduced in~\cite{jolliffe_2003}.
In~\cite{zou_hastie_tibshirani_2006}, a nonconvex regression-type optimization approach combined with LASSO penalty was used to approximately tackle the problem.
A nonconvex technique, locally solving difference-of-convex-functions programs, was presented in~\cite{sriperumbudur_torres_lanckriet_2007}.
Semidefinite programming (SDP) was used in~\cite{daspremont_chaoui_jordan_lanckriet_2007},~\cite{zhang_daspermont_elghaoui_2010}, while~\cite{daspermont_bach_elghaoui_2008} augmented the SDP approach with an extra greedy step that offers favorable optimality guarantees under certain sufficient conditions.
The authors of~\cite{moghaddam_weiss_avidan_2006b} considered greedy and branch-and-bound approaches.
Generalized power method techniques using convex programs were also used to approximately solve sparse PCA \cite{journee_nesterov_richtarik_sepulchre_2008}.
A sparse-adjusted deflation procedure was introduced in \cite{mackey_2009} and in \cite{Amini2008} optimality guarantees were shown for specific types of covariance matrices under thresholding and SDP relaxations.
Iterative thresholding was also considered in~\cite{Ma2011} in conjunction with certain guarantees while a truncated power method was presented in~\cite{Yuan2011}.

\IEEEpubidadjcol

In this present work, we prove that the sparse principal component of an $N\times N$ matrix ${\bf C}$ can be obtained in polynomial time under a new sufficient condition: when ${\bf C}$ can be written as a sum of a scaled identity matrix and a positive semidefinite update, i.e., ${\bf C}=\sigma{\bf I}_N+{\bf A}$, and the rank $D$ of the update ${\bf A}$ is not a function of the problem size.\footnote{If $\sigma=0$, then we simply have a constant-rank matrix ${\bf C}$.}
Under this condition, we show that sparse PCA is solvable with complexity ${\mathcal O}\left(N^{D+1}\right)$.
Our proof utilizes the auxiliary unit vector technique that we developed in~\cite{KP},~\cite{KL}.
This technique has been inspired by the work in~\cite{Mack:00}, which reappeared in~\cite{sweldens} and was used in~\cite{MLSD:00}.
It introduces an auxiliary unit vector that unlocks the constant-rank structure of a matrix (in this present work, matrix ${\bf A}$).
The constant-rank property along with the auxiliary vector enable us to scan a constant-dimensional space and identify a polynomial number of candidate vectors (i.e., candidate solutions to the original problem).
Interestingly, the optimal solution always lies among these candidates and a polynomial time search can always retrieve it.
As a result, we have applied the auxiliary unit vector technique to identify the polynomial solvability of certain optimization problems and provide polynomial-time algorithms that are directly implementable, fully parallelizable, and memory efficient~\cite{STBC}-\cite{panos}.

The rest of this paper is organized as follows.
In Section~\ref{section:problem}, we state the sparse PCA problem and indicate its ${\mathcal{NP}}$-hardness.
Then, in Section~\ref{section:parallel}, we follow the principles of the auxiliary unit vector technique to present a proof of the polynomial solvability of the sparse PCA problem under a constant-rank condition.
Moreover, we design a novel algorithm%
\footnote{Early versions of our algorithm appeared in~\cite{Asteris_thesis},~\cite{Asteris_ISIT}.}
which, for any sparsity value, computes the sparse principal component with complexity ${\mathcal O}\left(N^{D+1}\right)$.
Our algorithm is simply implementable, fully parallelizable, and memory efficient.
Especially for the case $D=2$, an alternative nonparallelizable version of our algorithm with complexity ${\mathcal O}\left(N^2\log N\right)$ is presented in Section~\ref{section:rank2}.%
\footnote{Alternative (non)parallelizable implementations of our algorithm for $D=3$ have been presented in~\cite{Asteris_thesis}.}
A few conclusions are drawn in Section~\ref{section:conclusions}.

\section{Problem Statement}
\label{section:problem}

We are interested in the computation of the real, unit-norm, and at most $K$-sparse principal component of a $N\times N$ matrix ${\bf C}$, i.e.,
\begin{equation}
\myvec{x}_{\text{opt}}\eqdef\!\!\!\!\!\!\argmax_{\begin{smallmatrix}{\bf x}\in{\mathbbm R}^N\\\|{\bf x}\|=1,\|{\bf x}\|_0\leq K\end{smallmatrix}}\!\!\!\!\!\!\left\{{\bf x}^T{\bf C}{\bf x}\right\}.
\label{initial_problem}
\end{equation}
Interestingly, when ${\bf C}$ can be decomposed as a constant-rank positive semidefinite update of the identity matrix, i.e.,
\begin{equation}
{\bf C}=\sigma{\bf I}_N+{\bf A}
\end{equation}
where $\sigma\in{\mathbbm R}$, ${\bf I}_N$ is the $N\times N$ identity matrix, and ${\bf A}$ is a positive semidefinite matrix with rank $D$, then the optimization~(\ref{initial_problem}) can always be rewritten as
\begin{equation}
\myvec{x}_{\text{opt}}=\!\!\!\!\!\!\argmax_{\begin{smallmatrix}{\bf x}\in{\mathbbm R}^N\\\|{\bf x}\|=1,\|{\bf x}\|_0\leq K\end{smallmatrix}}\!\!\!\!\!\!\left\{{\bf x}^T\left(\sigma{\bf I}_N+{\bf A}\right){\bf x}\right\}=\!\!\!\!\!\!\argmax_{\begin{smallmatrix}{\bf x}\in{\mathbbm R}^N\\\|{\bf x}\|=1,\|{\bf x}\|_0\leq K\end{smallmatrix}}\!\!\!\!\!\!\left\{{\bf x}^T{\bf A}{\bf x}\right\}.
\end{equation}
Since ${\bf A}$ is positive semidefinite and has rank $D$, it can be decomposed as
\begin{equation}
\mymatrix{A} = \mymatrix{V}\transp{\mymatrix{V}},
\end{equation}
where
\begin{equation}
\mymatrix{V}\stackrel{\bigtriangleup}{=}[\myvec{v}_1\;\myvec{v}_2\;\ldots\;\myvec{v}_D]
\end{equation}
is an $N\times D$ matrix, and problem~\eqref{initial_problem} can be written as
\begin{equation}
\myvec{x}_{\text{opt}}=\!\!\!\!\!\!\argmax_{\begin{smallmatrix}{\bf x}\in{\mathbbm R}^N\\\|{\bf x}\|=1,\|{\bf x}\|_0\leq K\end{smallmatrix}}\!\!\!\!\!\!\left\{{\bf x}^T{\bf V}{\bf V}^T{\bf x}\right\}=\!\!\!\!\!\!\argmax_{\begin{smallmatrix}{\bf x}\in{\mathbbm R}^N\\\|{\bf x}\|=1,\|{\bf x}\|_0\leq K\end{smallmatrix}}\!\!\!\!\!\!\left\|\transp{\mymatrix{V}}\myvec{x}\right\|.
\label{simpler_problem}
\end{equation}
For the optimization problem in~(\ref{simpler_problem}), we note that
\begin{equation}
\max_{\begin{smallmatrix}{\bf x}\in{\mathbbm R}^N\\\|{\bf x}\|=1,\|{\bf x}\|_0\leq K\end{smallmatrix}}\!\!\!\!\!\!\left\|{\bf V}^T{\bf x}\right\|
=
\max_{\begin{smallmatrix}\mathcal{I}\subseteq[N]\\|\mathcal{I}|=K\end{smallmatrix}}\max_{\begin{smallmatrix}{\bf x}\in{\mathbbm R}^K\\\|{\bf x}\|=1\end{smallmatrix}}\left\|{\bf V}_{{\mathcal I},:}^T{\bf x}\right\|
\label{eq:nested}
\end{equation}
where $\left[N\right]\eqdef\left\{1,2,\ldots,N\right\}$.
In~(\ref{eq:nested}), set ${\mathcal I}\subseteq[N]$ (which we call the {\it support}) consists of the indices of the $K$ potentially nonzero elements of ${\bf x}\in{\mathbbm R}^N$.
For a given support ${\mathcal I}$, the inner maximization is a $K$-dimensional principal-component problem, where ${\bf V}_{{\mathcal I},:}$ is the corresponding $K\times D$ submatrix of ${\bf V}$.
The solution to the inner maximization is denoted by
\begin{equation}
{\bf x}(\mathcal{I})\eqdef\argmax_{\begin{smallmatrix}{\bf x}\in{\mathbbm R}^K\\\|{\bf x}\|=1\end{smallmatrix}}\left\|{\bf V}_{{\mathcal I},:}^T{\bf x}\right\|
\label{eq:xI}
\end{equation}
and equals the principal left singular vector of ${\bf V}_{{\mathcal I},:}$.
Then, our optimization problem in~(\ref{eq:nested}) becomes
\begin{equation}
{\mathcal I}_\text{opt}\eqdef\argmax_{\begin{smallmatrix}\mathcal{I}\subseteq[N]\\|\mathcal{I}|=K\end{smallmatrix}}\left\{\sigma_{\max}\left({\bf V}_{{\mathcal I},:}\right)\right\}
\label{eq:maxsigma}
\end{equation}
where $\sigma_{\max}({\bf V})$ denotes the principal singular value of matrix ${\bf V}$.
That is, to solve our original problem in~(\ref{initial_problem}), according to~(\ref{eq:maxsigma}), we need to find the $K$-row submatrix of ${\bf V}$ whose principal singular value is the maximum one among all submatrices.
The indices that are contained in the optimal support ${\mathcal I}_\text{opt}$ that solves~(\ref{eq:maxsigma}) correspond to the nonzero loadings of the solution ${\bf x}_\text{opt}$ to~(\ref{initial_problem}).
Then, according to~(\ref{eq:xI}), the values of these nonzero loadings are directly computed by the left singular vector of ${\bf V}_{{\mathcal I}_\text{opt},:}$.

From the above discussion, it turns out that the hardness of the original problem in~(\ref{initial_problem}) comes from the identification of the optimal support ${\mathcal I}_\text{opt}$ in~(\ref{eq:maxsigma}).
To obtain the optimal support ${\mathcal I}_\text{opt}$, we could simply perform an exhaustive search among all $\binom{N}{K}$ possible supports ${\mathcal I}$ and compare them against the metric of interest in~(\ref{eq:maxsigma}).
However, if $K$ is not constant but grows with $N$, then such an approach has complexity that is exponential in $N$, indicating the ${\mathcal{NP}}$-hardness of~(\ref{initial_problem}), which was shown in~\cite{moghaddam_weiss_avidan_2006a}.
In this present work, we show that, if the rank $D$ of ${\bf A}$ is constant, then~(\ref{eq:maxsigma}) can be solved in time polynomial in $N$.
In fact, we develop an algorithm that has complexity ${\mathcal O}\left(N^{D+1}\right)$ and returns ${\mathcal O}\left(N^D\right)$ candidate supports, one of which is guaranteed to be the solution to~(\ref{eq:maxsigma}).
Then, by an exhaustive search among only these candidate supports, we identify the optimal support in~(\ref{eq:maxsigma}) and, hence, the sparse principal component of ${\bf A}$ and ${\bf C}$ with complexity polynomial in $N$, for any sparsity value $K$ between $1$ and $N$ (that is, even if $K$ grows with $N$).

\section{Computation of the Sparse Principal Component in Time ${\mathcal O}\left(N^{D+1}\right)$}
\label{section:parallel}

Prior to presenting the main result for the general rank-$D$ case, in the following subsection we provide insights as to why the sparse principal component of constant-rank matrices can be solved in polynomial time by first considering the trivial case $D=1$.

\subsection{Rank-$1$: A motivating example}
\label{subsection:rank_1_case}

In this case, $\mymatrix{A}$ has rank $1$ and $\mymatrix{V}=\myvec{v}\in{\mathbbm R}^N$.
For a given support ${\mathcal I}$, we have ${\bf V}_{{\mathcal I},:}={\bf v}_{\mathcal I}$.
Then, our optimization problem in~(\ref{eq:nested}) becomes
\begin{equation}
\max_{\begin{smallmatrix}\mathcal{I}\subseteq[N]\\|\mathcal{I}|=K\end{smallmatrix}}\max_{\begin{smallmatrix}{\bf x}\in{\mathbbm R}^K\\\|{\bf x}\|=1\end{smallmatrix}}\left|{\bf v}_{\mathcal I}^T{\bf x}\right|
\label{eq:rank1}
\end{equation}
where, for any given support ${\mathcal I}$, the corresponding vector in~(\ref{eq:xI}) is
\begin{equation}
{\bf x}(\mathcal{I})=\argmax_{\begin{smallmatrix}{\bf x}\in{\mathbbm R}^K\\\|{\bf x}\|=1\end{smallmatrix}}\left|{\bf v}_{\mathcal I}^T{\bf x}\right|=\frac{{\bf v}_{\mathcal I}}{\left\|{\bf v}_{\mathcal I}\right\|}.
\end{equation}
Therefore,~(\ref{eq:nested}) becomes
\begin{equation}
\max_{\begin{smallmatrix}\mathcal{I}\subseteq[N]\\|\mathcal{I}|=K\end{smallmatrix}}\left|{\bf v}_{\mathcal I}^T\frac{{\bf v}_{\mathcal I}}{\left\|{\bf v}_{\mathcal I}\right\|}\right|=\max_{\begin{smallmatrix}\mathcal{I}\subseteq[N]\\|\mathcal{I}|=K\end{smallmatrix}}\left\|{\bf v}_{\mathcal I}\right\|
\end{equation}
and the optimal support is
\begin{equation}
{\mathcal I}_\text{opt}=\argmax_{\begin{smallmatrix}\mathcal{I}\subseteq[N]\\|\mathcal{I}|=K\end{smallmatrix}}\left\|{\bf v}_{\mathcal I}\right\|=\argmax_{\begin{smallmatrix}\mathcal{I}\subseteq[N]\\|\mathcal{I}|=K\end{smallmatrix}}\sum_{n\in{\mathcal I}}v_n^2.
\end{equation}
That is, to determine the solution to~(\ref{eq:maxsigma}), we only need to compare the elements of $|{\bf v}|$ and select the $K$ largest ones.
Then, their indices are the elements of ${\mathcal I}_\text{opt}$.

The above observation, although simple, turns out to be critical for the developments that follow.
Hence, to simplify the presentation, we define function $\text{\it top}_k$ which is parameterized in an integer $k$, has as input a vector ${\bf u}$ of length $N\geq k$, and returns the indices of the $k$ largest values in $|{\bf u}|$:
\begin{equation}
\text{top}_k({\bf u})\eqdef\argmax_{\begin{smallmatrix}\mathcal{I}\subseteq[N]\\|\mathcal{I}|=k\end{smallmatrix}}\left\|{\bf u}_{\mathcal I}\right\|.
\end{equation}
Function $\text{top}_k({\bf u})$ operates by selecting the indices of the $k$ largest values among $|u_1|$, $|u_2|$, $\ldots$, $|u_N|$.
Its complexity is ${\mathcal O}\left(N\right)$~\cite{cormen}.

We conclude this subsection by noting that, if $D=1$, then the optimal support in~(\ref{eq:maxsigma}) is
\begin{equation}
{\mathcal I}_\text{opt}=\text{top}_K({\bf v})
\label{eq:ID1}
\end{equation}
and is computed with linear complexity.

\subsection{Rank-$D$: Utilizing the auxiliary unit vector technique}

We consider now the case where $\mymatrix{A}$ has rank $D\geq1$ and, hence, $\mymatrix{V}$ is an $N\times D$ matrix.
Without loss of generality (w.l.o.g.), we assume that each row of $\mymatrix{V}$ has at least one nonzero element, i.e., $\mymatrix{V}_{n,1:D} \neq \myvec{0}_{1\times D}$, $\forall\;n\in[N]$.
Indeed, as explained in~\cite{KL}, if there exists an index $n\in[N]$ such that $\mymatrix{V}_{n,:}=\myvec{0}$, then, independently of the value of the corresponding element $x_n$ of $\myvec{x}$, the contribution of this row to the value of $\left\|\transp{\mymatrix{V}}\myvec{x}\right\|$ in~(\ref{simpler_problem}) will be zero.
Hence, there is no point in ``spending'' in $x_n$ a weight that could be distributed to other elements of $\myvec{x}$; we can ignore the $n$th row of $\mymatrix{V}$, replace $\mymatrix{V}$ by $\mymatrix{V}_{[N]-\{n\},:}$, and, hence, reduce the problem size from $N$ to $N-1$.
In the final solution $\myvec{x}_{\text{opt}}$, $x_n$ will be set to zero.

In our subsequent developments, we rely on the auxiliary unit vector technique that was introduced in~\cite{KP} for matrices of rank $D=2$ and generalized in~\cite{KL} for matrices of rank $D\geq2$.
This technique utilizes an auxiliary vector ${\bf c}$ to generate the subspace spanned by the columns of ${\bf V}$ and result in a rank-$1$ problem for each value of ${\bf c}$.
Interestingly, for several problems, the number of different solutions that we obtain as ${\bf c}$ scans the unit-radius hypersphere has polynomial size.
If the rank-$1$ problem for each value of ${\bf c}$ is polynomially solvable (as, for example, in the optimization problem of this work, as indicated in Subsection~\ref{subsection:rank_1_case}), then the optimal solution is obtained with overall polynomial complexity.
In a few words, the auxiliary unit vector technique of~\cite{KP},~\cite{KL} is a fully parallelizable and memory efficient technique that translates $D$-dimensional problems into a polynomial collection of rank-$1$ problems among which one results in the overall optimal solution.

For our sparse-principal-component problem, the auxiliary unit vector technique works as follows.
Consider a unit vector ${\bf c}\in{\mathbbm R}^D$.
By Cauchy-Schwartz Inequality, for any ${\bf a}\in{\mathbbm R}^D$,
\begin{equation}
\left|{\bf a}^T{\bf c}\right|\leq\left\|{\bf a}\right\|\left\|{\bf c}\right\|=\left\|{\bf a}\right\|
\end{equation}
with equality if and only if ${\bf c}$ is collinear to ${\bf a}$.
Then,
\begin{equation}
\max_{{\bf c}\in{\mathbbm R}^D,\left\|{\bf c}\right\|=1}\left|{\bf a}^T{\bf c}\right|=\left\|{\bf a}\right\|.
\label{eq:ca}
\end{equation}
Using~(\ref{eq:ca}), our optimization problem in~(\ref{eq:nested}) becomes
\begin{align}
\hspace{-.1cm}\max_{\begin{smallmatrix}\mathcal{I}\subseteq[N]\\|\mathcal{I}|=K\end{smallmatrix}}\max_{\begin{smallmatrix}{\bf x}\in{\mathbbm R}^K\\\|{\bf x}\|=1\end{smallmatrix}}\left\|{\bf V}_{{\mathcal I},:}^T{\bf x}\right\|&=\hspace{-.1cm}\max_{\begin{smallmatrix}\mathcal{I}\subseteq[N]\\|\mathcal{I}|=K\end{smallmatrix}}\max_{\begin{smallmatrix}{\bf x}\in{\mathbbm R}^K\\\|{\bf x}\|=1\end{smallmatrix}}\max_{\begin{smallmatrix}{\bf c}\in{\mathbbm R}^D\\\left\|{\bf c}\right\|=1\end{smallmatrix}}\left|{\bf x}^T{\bf V}_{{\mathcal I},:}{\bf c}\right|\nonumber\\
&=\hspace{-.1cm}\max_{\begin{smallmatrix}{\bf c}\in{\mathbbm R}^D\\\left\|{\bf c}\right\|=1\end{smallmatrix}}\max_{\begin{smallmatrix}\mathcal{I}\subseteq[N]\\|\mathcal{I}|=K\end{smallmatrix}}\max_{\begin{smallmatrix}{\bf x}\in{\mathbbm R}^K\\\|{\bf x}\|=1\end{smallmatrix}}\left|{\bf x}^T{\bf u}_{\mathcal I}\!\left({\bf c}\right)\right|
\label{eq:maxmaxmax}
\end{align}
where
\begin{equation}
{\bf u}\!\left({\bf c}\right)\eqdef{\bf V}{\bf c}.
\label{eq:uc}
\end{equation}
The rightmost equality in~(\ref{eq:maxmaxmax}) is obtained by interchanging the maximizations.
This is a critical step of the auxiliary unit vector technique.
It unlocks the constant-rank structure of ${\bf V}$ and allows us to consider a simple rank-$1$ problem for each value of ${\bf c}$.
Indeed, for each ${\bf c}\in{\mathbbm R}^D$, the inner double maximization problem
\begin{equation}
\max_{\begin{smallmatrix}\mathcal{I}\subseteq[N]\\|\mathcal{I}|=K\end{smallmatrix}}\max_{\begin{smallmatrix}{\bf x}\in{\mathbbm R}^K\\\|{\bf x}\|=1\end{smallmatrix}}\left|{\bf u}_{\mathcal I}\!\left({\bf c}\right)^T{\bf x}\right|
\end{equation}
is equivalent to the rank-$1$ optimization problem in~(\ref{eq:rank1}) that, according to~(\ref{eq:ID1}), results in the optimal support (for fixed ${\bf c}$)
\begin{equation}
{\mathcal I}({\bf c})\eqdef\text{top}_K({\bf u}\!\left({\bf c}\right))
\label{eq:Ic}
\end{equation}
which is obtained with complexity ${\mathcal O}(N)$.
Then, according to~(\ref{eq:maxmaxmax}), the solution to our original problem in~(\ref{eq:maxsigma}) is met by collecting all possible supports ${\mathcal I}({\bf c})$ as ${\bf c}$ scans the unit-radius $D$-dimensional hypersphere.
That is, ${\mathcal I}_\text{opt}$ in~(\ref{eq:maxsigma}) belongs to
\begin{equation}
{\mathcal S}\eqdef\!\!\!\!\!\!\bigcup_{{\bf c}\in{\mathbbm R}^D,\left\|{\bf c}\right\|=1}\!\!\!\!\!\!{\mathcal I}({\bf c}).
\label{eq:S}
\end{equation}

Set ${\mathcal S}$ contains candidate supports ${\mathcal I}\subseteq[N]$ one of which is the solution to our original optimization problem.
If ${\mathcal S}$ was available, then one would have to compare the elements of ${\mathcal S}$ against the metric of interest in~(\ref{eq:maxsigma}) to obtain the optimal support ${\mathcal I}_\text{opt}$.
Therefore, the size of ${\mathcal S}$ and the complexity to build ${\mathcal S}$ determine the overall complexity to solve~(\ref{eq:maxsigma}).
Our major contribution in this work is that we show that the cardinality of ${\mathcal S}$ is upper bounded by
\begin{equation}
\left|{\mathcal S}\right|\leq2^{D-1}\binom{D}{\left\lfloor\frac{D}{2}\right\rfloor}\binom{N}{D}={\mathcal O}\left(N^D\right)
\end{equation}
and develop an algorithm to build ${\mathcal S}$ with complexity ${\mathcal O}\left(N^{D+1}\right)$.
After ${\mathcal S}$ is constructed, each element (support) ${\mathcal I}$ of it is mapped to the principal singular value of the $K\times D$ matrix ${\bf V}_{{\mathcal I},:}$ with complexity ${\mathcal O}\left(KD^2\right)={\mathcal O}\left(K\right)$, since $D$ is constant.
Finally, all computed singular values are compared with each other to obtain the optimal support ${\mathcal I}_\text{opt}$ in~(\ref{eq:maxsigma}).
Then, the solution to our original problem in~(\ref{initial_problem}) is the principal left singular vector of the $K\times D$ matrix ${\bf V}_{{\mathcal I}_\text{opt},:}$, computed with complexity ${\mathcal O}\left(KD^2\right)={\mathcal O}\left(K\right)$.
Therefore, we compute the optimal support ${\mathcal I}_\text{opt}$ and the sparse principal component of a rank-$D$ matrix with complexity ${\mathcal O}\left(N^{D+1}+N^DK\right)={\mathcal O}\left(N^{D+1}\right)$.

A constructive proof is presented in detail in the next three subsections.
To give some insight of the proof, we begin with the simple case $D=2$.
Then, we generalize our proof for the case of any arbitrary $D$.

\subsection{Rank-$2$: A simple instance of our proof}
\label{subsection:rank2}

If $D=2$, then ${\bf V}$ has size $N\times2$ and the auxiliary vector ${\bf c}$ is a length-$2$, unit vector that, as in~\cite{KP}, can be parameterized in an auxiliary angle $\phi\in\left(-\frac{\pi}{2},\frac{\pi}{2}\right]$.
That is,
\begin{equation}
\myvec{c}(\phi)\stackrel{\bigtriangleup}{=} \begin{bmatrix}
\sin\phi\\
\cos\phi
\end{bmatrix},\;\;\;\phi\in\Phi\eqdef\left(-\frac{\pi}{2},\frac{\pi}{2}\right].
\end{equation}
Hence, ${\bf c}(\phi)$ lies on the unit-radius semicircle.%
\footnote{We ignore the other semicircle because any pair of angles $\phi_1$ and $\phi_2$ with difference $\pi$ results in opposite vectors ${\bf c}(\phi_1)=-{\bf c}(\phi_2)$ which, however, are equivalent with respect to the optimization metric in~(\ref{eq:maxmaxmax}) and produce the same support ${\mathcal I}\left({\bf c}(\phi_1)\right)={\mathcal I}\left({\bf c}(\phi_2)\right)$ in~(\ref{eq:Ic}).}
Then, the candidate set in~(\ref{eq:S}) is re-expressed as
\begin{equation}
{\mathcal S}=\bigcup_{\phi\in\Phi}{\mathcal I}(\phi)
\end{equation}
where, according to~(\ref{eq:Ic}),
\begin{equation}
{\mathcal I}(\phi)\eqdef\text{top}_K({\bf u}(\phi))
\end{equation}
and, according to~(\ref{eq:uc}),
\begin{equation}
{\bf u}\!\left(\phi\right)\eqdef{\bf V}{\bf c}\!\left(\phi\right).
\end{equation}
That is, for any given $\phi\in\Phi$, the corresponding support ${\mathcal I}(\phi)$ is obtained with complexity ${\mathcal O}(N)$ by selecting the indices of the $K$ absolutely largest elements of $\myvec{u}(\phi)$.

However, why should $\phi$ simplify the computation of a solution?
The intuition behind the auxiliary unit vector technique is that every element of $\myvec{u}(\phi)$ is actually a continuous function of $\phi$, i.e., a curve (or $1$-manifold) in $\phi$, since
\begin{equation}
{\bf u}(\phi)=\mymatrix{V}\myvec{c}(\phi)=\left[\begin{array}{c}
V_{1,1}\sin{\phi}+V_{1, 2}\cos{\phi}\\
V_{2,1}\sin{\phi}+V_{2, 2}\cos{\phi}\\
\vdots\\
V_{N,1}\sin{\phi}+V_{N, 2}\cos{\phi}\end{array}\right].
\label{rank2intro:Vc}
\end{equation}
Hence, the support that corresponds to the $K$ absolutely largest elements of $\myvec{u}(\phi)$ at a given point $\phi$ is a function of $\phi$. 
Due to the continuity of the curves and the discrete nature of the support, we expect that the support ${\mathcal I}(\phi)$ will retain the same elements in an area around $\phi$.
Therefore, we expect the formation of intervals in $\Phi$, within which the indices of the $K$ absolutely largest elements of ${\bf u}(\phi)$ remain unaltered.
A support ${\mathcal I}$ might change only if the sorting of the amplitudes of two elements in ${\bf u}(\phi)$, say $|u_i(\phi)|$ and $|u_j(\phi)|$, changes.
This occurs at points $\phi$ where $|u_i(\phi)|=|u_j(\phi)|$, that is, points where two curves intersect.
Finding all these {\it intersection points} is sufficient to determine intervals and construct all possible candidate supports ${\mathcal I}$.
Among all candidate supports, lies the support that corresponds to the optimal $K$-sparse principal component.
Exhaustively checking the supports ${\mathcal I}$ of all intervals suffices to retrieve the optimal one.
The number of these intervals is exactly equal to number of possible intersections among the amplitudes of $\left|{\bf u}(\phi)\right|$, which is exactly equal to $2\binom{N}{2}=\mathcal{O}\left(N^2\right)$, counting all possible combinations of element pairs.

\begin{figure}[t!]
\centerline{\epsfig{file=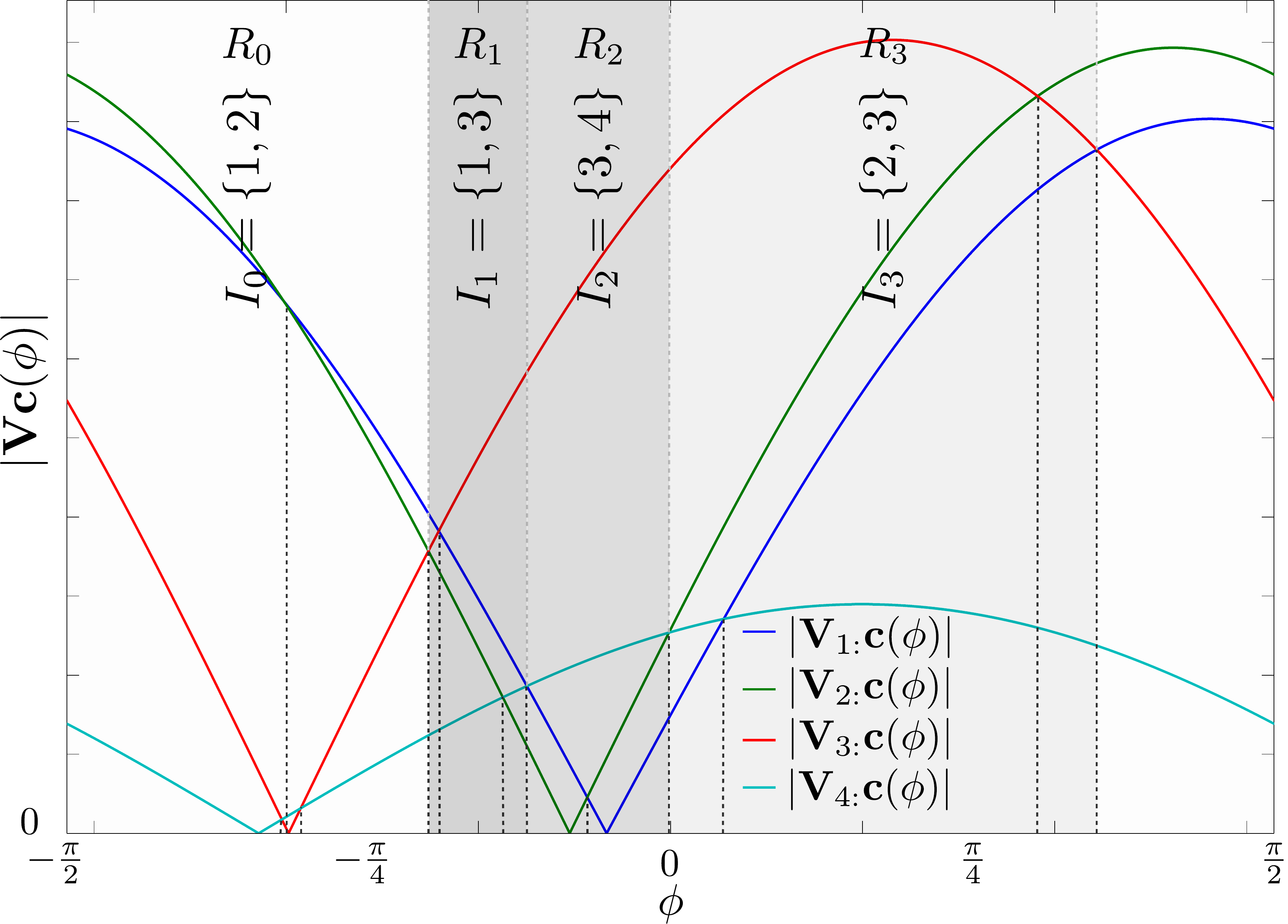,width=\myfigurewidth}}
\caption{Partition of $\Phi$ into $2\binom{4}{2}=12$ intervals and $4$ regions (for sparsity $K=2$), generated by the rows of an arbitrary $4\times2$ matrix ${\bf V}$.}
\label{rank2intro:fig:V_Rank2_Cells_and_Regions}
\end{figure}

Before we proceed, in Fig.~\ref{rank2intro:fig:V_Rank2_Cells_and_Regions}, we illustrate the interval partition of $\Phi$ for an arbitrary $4\times2$ matrix ${\bf V}$ (i.e., $N=4$).
We plot the $4$ curves that originate from the $4$ rows of $\left|{\bf u}(\phi)\right|=\left|\mymatrix{V}\myvec{c}(\phi)\right|$ and observe the intervals that are formed, within which the sorting of the curves does not change.
The borders of the intervals are denoted by vertical dashed lines at points of curve intersections.
Our approach creates $2\binom{4}{2}=12$ intervals which exceeds the total number of possible supports, however this is not true for greater values of $N$.
In addition, for sparsity $K=2$, we observe that $\Phi$ is partitioned into {\it regions} (sets of adjacent intervals); within each region $R_i$, although the sorting changes, the set of $K$ largest curves does not change.
For example, in Fig.~\ref{rank2intro:fig:V_Rank2_Cells_and_Regions}, we identify the regions $R_0$, $R_1$, $R_2$, and $R_3$ where the candidate support remains fixed.
These regions are an interesting feature that might further decrease the number of intervals we need to check.
We exploit this feature in the serial implementation of our algorithm in Section~\ref{section:rank2}.

\subsection{Rank-$2$: Algorithmic developments and complexity}
\label{subsection:rank2algo}

Our goal is the construction of all possible candidate $K$-sparse vectors, determined by the support ${\mathcal I}$ of each interval in $\Phi$.
This is a two-step process.
First, we identify interval borders and, then, we determine the supports associated with these intervals.

{\it Algorithmic Steps:}
We first determine all possible intersections of curve pairs in $|{\bf u}(\phi)|$.
Any pair $\{i,j\}$ of distinct elements in $|{\bf u}(\phi)|$ is associated with two intersections: $u_i(\phi)=u_j(\phi)$ and $u_i(\phi)=-u_j(\phi)$. 
Solving these two equations with respect to $\phi$ determines two points
\begin{equation}
\hat\phi=\mp\tan^{-1}\left(\frac{V_{i,2}-V_{j,2}}{V_{i,1}-V_{j,1}}\right)\in\Phi
\label{eq:hatphi}
\end{equation}
where a new support of indices of the $K$-largest values of $|{\bf u}(\phi)|$ might occur.
We note that, when $\phi$ varies in $\Phi$ (from $-\frac{\pi}{2}$ to $\frac{\pi}{2}$), changes of the support ${\mathcal I}$ may occur only over intersection points $\hat\phi$ given in~(\ref{eq:hatphi}), for any $i,j\in[N]$ with $i\neq j$.

Next, we observe that, at an intersection point $\hat\phi$, the intersecting curves are equal, i.e., $|u_i(\hat\phi)|=|u_j(\hat\phi)|$, while their relative order changes over $\hat{\phi}$.
That is, if $|u_i(\phi)| > |u_j(\phi)|$ in the interval immediately preceding $\hat{\phi}$, then $|u_i(\phi)|<|u_j(\phi)|$ in the interval immediately following $\hat{\phi}$, and vice versa.
Exactly at the point of intersection, we can determine the order of the curves in $|\mathbf{u}(\hat{\phi})|$.
\begin{itemize}
\item
If both or none of the $i$th and $j$th curves are included in the set of $K$ largest curves at $\hat{\phi}$, then none of the two curves leaves or joins, respectively, that set at $\hat{\phi}$, despite the change in their relative order.
Assuming that no other pair of curves intersect exactly at $\hat{\phi}$,%
\footnote{Recall that the support ${\mathcal I}$ can change only at an intersection point.
If a second pair of curves, say the $l$th and $k$th ones, intersect exactly at $\hat{\phi}$ causing the support ${\mathcal I}$ to change over $\hat{\phi}$, then the correct support sets for the two incident intervals will be determined when the intersections of the $l$th and $k$th curves are examined.}
the two intervals incident to $\hat{\phi}$ are associated with the same support ${\mathcal I}$ which can be determined exactly at the intersection point $\hat{\phi}$.%
\footnote{In fact, since the support does not change at $\hat{\phi}$, there exists another intersection point where the same support will be collected.
Hence, the support ${\mathcal I}$ at $\hat{\phi}$ may safely be ignored.}
\item
If that is not the case, i.e., if the $i$th and $j$th curves occupy the $K$th and $(K+1)$th order at $\hat{\phi}$, then the change in their relative order affects the support ${\mathcal I}$: one of the two curves leaves and the other one joins the set of $K$ largest curves at $\hat{\phi}$.
The support sets associated with the two adjacent intervals differ only in one element (one contains index $i$ and the other contains index $j$ instead), while the remaining $K-1$ common indices correspond to the $K-1$ largest curves at the intersection point $\hat{\phi}$.%
\footnote{Since any interval in $\Phi$ follows an intersection point, we actually need to evaluate the support ${\mathcal I}$ of only the interval that follows the intersection point $\hat{\phi}$.
The other interval will be examined at the intersection point that precedes it.
To identify which of the two indices $i$ and $j$ is contained in the support of the interval that follows the intersection point $\hat\phi$,
we can visit the ``rightmost'' point of $\Phi$, that is, $\frac{\pi}{2}$.
There, due to the continuity of $u_i(\phi)$ and $\pm u_j(\phi)$, the relative order of $|u_i(\phi)|$ and $|u_j(\phi)|$ within the interval that follows $\hat\phi$ will be the identical or opposite relative order of $u_i\left(\frac{\pi}{2}\right)$ and $\pm u_j\left(\frac{\pi}{2}\right)$, depending on whether $u_i(\hat\phi)$ is positive or negative, respectively.}
\end{itemize}
We have fully described a way to calculate the (at most) two support sets associated with the two intervals incident to one intersection point.
Since all intervals are incident to at least one intersection point, it suffices to consider all intersection points, that is, all $\binom{N}{2}$ pairwise combinations of elements in $|{\bf u}(\phi)|$, to collect the corresponding support sets.

{\it Computational Complexity:}
A single intersection point can be computed in time $\mathcal{O}(1)$.
At an intersection point $\hat\phi$, the $K$th-order element of $|\mathbf{u}(\phi)|$ (and the $K-1$ elements larger than that) can be determined in time $\mathcal{O}(N)$ which equals the construction time of the (at most) two support sets associated with the intervals incident to $\hat\phi$.
Collecting all candidate supports ${\mathcal{I}}$ requires examining all $2{N \choose 2}$ intersection points, implying a total construction cost of $2{N \choose 2}\times{\mathcal O}\left(N\right)=\mathcal{O}\left(N^3\right)$.
Since we obtain (at most) two supports for any of the $2{N \choose 2}$ intersection points, the size of the candidate support set ${\mathcal S}$ is $|{\mathcal S}|\leq4{N \choose 2}$.

\subsection{Rank-$D$: A generalized proof}
\label{subsection:rankD}

\begin{figure*}[!b]
\addtocounter{equation}{4}
\hrulefill
\begin{equation}
{\bf u}({\boldsymbol\phi})
=\mymatrix{V} \myvec{c}(\myvecsymb{\phi})
=\left[\begin{array}{c}
{\bf V}_{1,:}{\bf c}({\boldsymbol\phi})\\
{\bf V}_{2,:}{\bf c}({\boldsymbol\phi})\\
\vdots \\
{\bf V}_{N,:}{\bf c}({\boldsymbol\phi})
\end{array}\right]
=\left[\begin{array}{c}
V_{1,1}\sin{\phi_1}+\sum_{d=2}^{D-1}{V_{1,d}\prod_{i=1}^{d-1}{\cos{\phi_{i}}}\sin{\phi_{d}}+V_{1,D}\prod_{i=1}^{D-1}{\cos{\phi_{i}}}}\\
V_{2,1}\sin{\phi_1}+\sum_{d=2}^{D-1}{V_{2,d}\prod_{i=1}^{d-1}{\cos{\phi_{i}}}\sin{\phi_{d}}+V_{2,D}\prod_{i=1}^{D-1}{\cos{\phi_{i}}}}\\
\vdots \\
V_{N,1}\sin{\phi_1}+\sum_{d=2}^{D-1}{V_{N,d}\prod_{i=1}^{d-1}{\cos{\phi_{i}}}\sin{\phi_{d}}+V_{N,D}\prod_{i=1}^{D-1}{\cos{\phi_{i}}}}
\end{array}\right]
\label{Vc}
\end{equation}
\addtocounter{equation}{-5}
\end{figure*}

In the general case, $\mymatrix{V}$ is a $N\times D$ matrix.
In this subsection, we present our main result where we prove that the problem of identifying the $K$-sparse principal component of a rank-$D$ matrix is solvable with complexity ${\mathcal O}\left(N^{D+1}\right)$.
The statement is true for any value of $K$ (that is, even if $K$ is a function of $N$).
The rest of this subsection contains a constructive proof of this statement.

Since ${\bf V}$ has size $N\times D$, the auxiliary vector ${\bf c}$ is a length-$D$, unit vector.
We begin our constructive proof by introducing the auxiliary-angle vector ${\boldsymbol\phi}\in\Phi^{D-1}$ and parameterizing ${\bf c}$, as in~\cite{KL}, according to
\begin{equation}
\myvec{c}(\myvecsymb{\phi})\eqdef\begin{bmatrix}
\sin\phi_1\\
\cos\phi_1\sin\phi_2\\
\cos\phi_1\cos\phi_2\sin\phi_3\\
\vdots\\ 
\cos\phi_1\cos\phi_2\hdots\sin\phi_{D-1}\\
\cos\phi_1\cos\phi_2\hdots\cos\phi_{D-1}                                                         
\end{bmatrix}.
\end{equation}
Hence, ${\bf c}({\boldsymbol\phi})$ lies on the unit-radius semihypersphere.%
\footnote{As in the rank-$2$ case, we ignore the other semihypersphere because any pair of vectors ${\boldsymbol\phi}$ and $\tilde{\boldsymbol\phi}$ whose first elements $\phi_1$ and $\tilde{\phi}_1$, respectively, have difference $\pi$ results in opposite vectors ${\bf c}(\boldsymbol\phi)=-{\bf c}(\tilde{\boldsymbol\phi})$ which, however, are equivalent with respect to the optimization metric in~(\ref{eq:maxmaxmax}) and produce the same support ${\mathcal I}\left({\bf c}({\boldsymbol\phi})\right)={\mathcal I}({\bf c}(\tilde{\boldsymbol\phi}))$ in~(\ref{eq:Ic}).}
Then, the candidate set in~(\ref{eq:S}) is re-expressed as
\begin{equation}
{\mathcal S}=\bigcup_{{\boldsymbol\phi}\in\Phi^{D-1}}{\mathcal I}({\boldsymbol\phi})
\label{eq:SD}
\end{equation}
where, according to~(\ref{eq:Ic}),
\begin{equation}
{\mathcal I}({\boldsymbol\phi})\eqdef\text{top}_K({\bf u}({\boldsymbol\phi}))
\end{equation}
and, according to~(\ref{eq:uc}),
\begin{equation}
{\bf u}\!\left({\boldsymbol\phi}\right)\eqdef{\bf V}{\bf c}\!\left({\boldsymbol\phi}\right).
\end{equation}
That is, for any given ${\boldsymbol\phi}\in\Phi^{D-1}$, the corresponding support ${\mathcal I}({\boldsymbol\phi})$ is obtained with complexity ${\mathcal O}(N)$ by selecting the indices of the $K$ absolutely largest elements of $\myvec{u}({\boldsymbol\phi})$.

\begin{figure}[t!]
\centerline{\epsfig{file=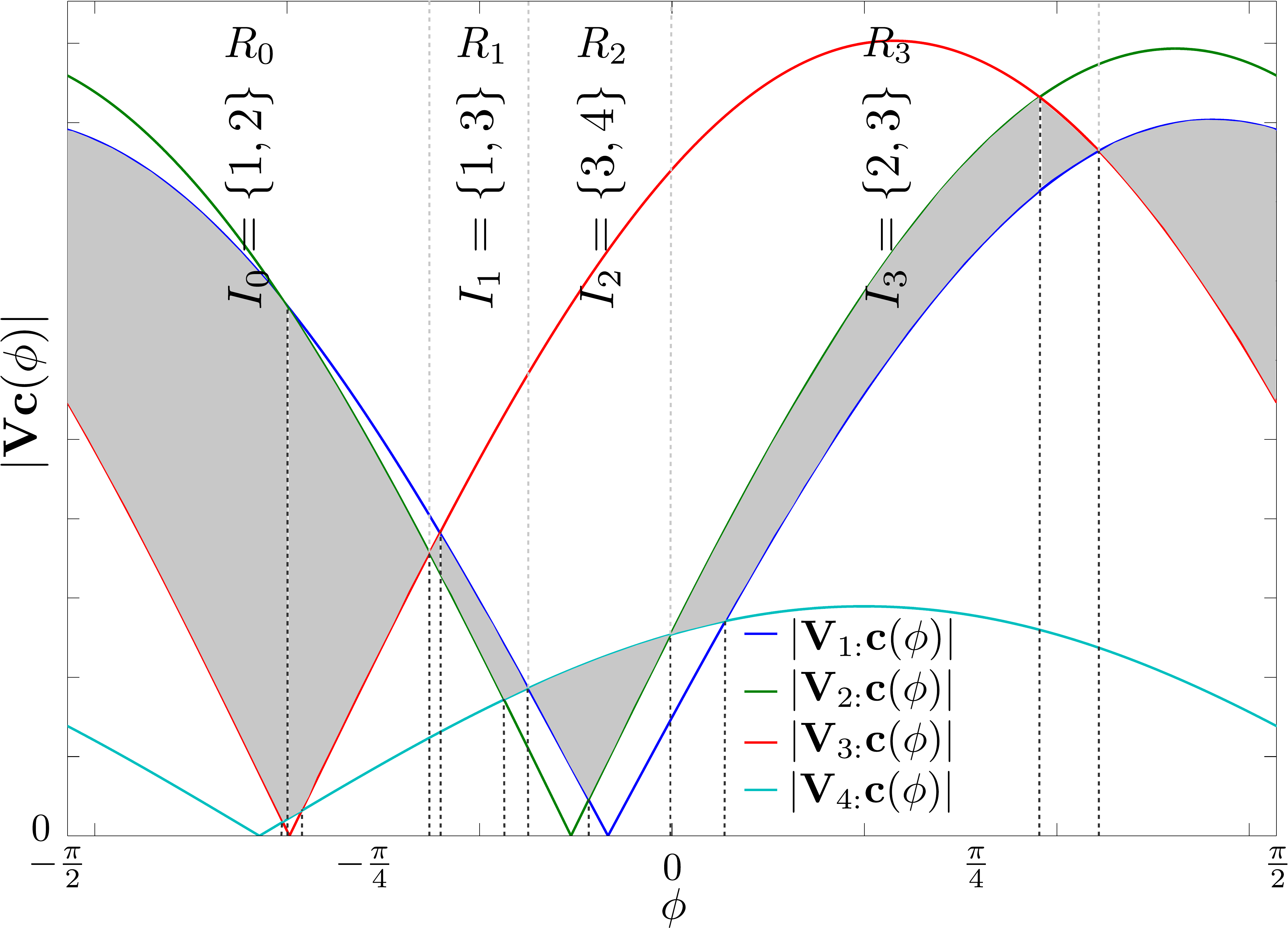,width=\myfigurewidth}}
\caption{Cells generated by the rows of an arbitrary $4\times2$ matrix ${\bf V}$.
The shaded cells are the normal ones for sparsity $K=2$.}
\label{fig:normalcells}
\end{figure}

To gain some intuition into the purpose of inserting the auxiliary-angle vector $\myvecsymb{\phi}$, notice that every element of ${\bf u}({\boldsymbol\phi})$ in~(\ref{Vc})
\addtocounter{equation}{1}%
is actually a continuous function of $\myvecsymb{\phi}$ and so are the elements of $|{\bf u}(\myvecsymb{\phi})|$.
That is, each element of $|{\bf u}(\myvecsymb{\phi})|$ is a hypersurface (or $(D-1)$-manifold) in the $D$-dimensional space $\Phi^{D-1}\times[0,\infty)$.
When we sort the $N$ elements of $|\myvec{u}(\myvecsymb{\phi})|$ at a given point $\myvecsymb{\phi}$, we actually sort the $N$ hypersurfaces at point $\myvecsymb{\phi}$.
The key observation in our algorithm is that, due to their continuity, the hypersurfaces will retain their sorting in an area ``around'' $\myvecsymb{\phi}$.
This implies the partition of $\Phi^{D-1}\times[0,\infty)$ into cells ${\mathcal C}_1,{\mathcal C}_2,\ldots,$ each of which (say, cell ${\mathcal C}$) is associated with a single set ${\mathcal I}^+({\mathcal C})\subseteq[N]$ of indices of hypersurfaces that lie above ${\mathcal C}$ and a single set ${\mathcal I}^-({\mathcal C})=[N]-{\mathcal I}^+({\mathcal C})$ of indices of hypersurfaces that lie below it.
Moreover, each cell ${\mathcal C}$ contains at least one vertex (that is, intersection of $D$ hypersurfaces).
Finally, for any $\myvecsymb{\phi}\in\Phi$, there is a unique cell ${\mathcal C}\subset\Phi^{D-1}\times[0,\infty)$, called ``normal,'' which contains uncountably many points in $\{{\boldsymbol\phi}\}\times[0,\infty)$ and is associated with a single index-set ${\mathcal I}^+({\mathcal C})$ of cardinality $K$ (that is, exactly $K$ hypersurfaces lie above ${\mathcal C}$).
In fact, the indices of these $K$ hypersurfaces (that is, the elements of ${\mathcal I}^+({\mathcal C})$) are the elements of support ${\mathcal I}({\boldsymbol\phi})$.
Although our discussion refers to the general-$D$ case, for illustrative purposes we consider again the case $D=2$ and, in Fig.~\ref{fig:normalcells}, we revisit the example that we presented in Subsection~\ref{subsection:rank2}.
The normal cells that are created by the $N=4$ curves are the shaded ones.
These cells carry the property that lie below exactly $K=2$ curves.
We observe that there is a one-to-one correspondence between normal cells and regions $R_i$.

According to~(\ref{eq:SD}) and the above observations, we need to determine the index-set ${\mathcal I}^+({\mathcal C})$ of every normal cell ${\mathcal C}$ in the partition.
If we collect all such index-sets, then we have constructed ${\mathcal S}$ in~(\ref{eq:SD}).
This will be achieved if, instead, we identify all cells in $\Phi^{D-1}\times[0,\infty)$ and, for each cell, determine the $K$ largest hypersurfaces that lie above an arbitrary point of it.
The latter will return the desired index-set ${\mathcal I}^+({\mathcal C})$ if the cell is normal.
In Fig.~\ref{fig:normalcells}, we observe that, for each normal cell, the indices of the $K=2$ largest curves that lie above it can be computed at the leftmost vertex of it (we can ignore the leftmost normal cell because it produces the same indices with the rightmost one).
In the following, we identify all cells in the partition and compute a size-$K$ support ${\mathcal I}$ for each such cell.
This way, we obtain the index-set of any normal cell, among which one is the optimal support ${\mathcal I}$ in~(\ref{eq:maxsigma}).

Since each cell contains at least one vertex, we only need to find all vertices in the partition and determine ${\mathcal I}^+$ for all neighboring cells.
Recall that a vertex is an intersection of $D$ hypersurfaces.
Consider $D$ arbitrary hypersurfaces; say, for example, $|u_1({\boldsymbol\phi})|$, $|u_2({\boldsymbol\phi})|$, $\ldots$, $|u_D({\boldsymbol\phi})|$.
Their intersection satisfies
$|u_1({\boldsymbol\phi})|=|u_2({\boldsymbol\phi})|=\ldots=|u_D({\boldsymbol\phi})|$ and is computed by solving the system of equations
\begin{equation}
\left\{\begin{array}{c}
u_1({\boldsymbol\phi})\pm u_2({\boldsymbol\phi})=0\\
u_1({\boldsymbol\phi})\pm u_3({\boldsymbol\phi})=0\\
\vdots\\
u_1({\boldsymbol\phi})\pm u_D({\boldsymbol\phi})=0\end{array}\right\}
\end{equation}
or, equivalently,
\begin{equation}
\left[\begin{array}{c}
{\bf V}_{1,:}\pm{\bf V}_{2,:}\\
{\bf V}_{1,:}\pm{\bf V}_{3,:}\\
\vdots\\
{\bf V}_{1,:}\pm{\bf V}_{D,:}\end{array}\right]{\bf c}({\boldsymbol\phi})={\bf 0}.
\label{eq:VVc}
\end{equation}
For any sign combination, the solution to the latter consists of the spherical coordinates of the unit vector in the null space of the $(D-1)\times D$ leftmost matrix.%
\footnote{If the $(D-1)\times D$ matrix is full-rank, then its null space has rank $1$ and ${\bf c}({\boldsymbol\phi})$ is uniquely determined (within a sign ambiguity which, however, does not affect the final decision on the index-set).
If, instead, the $(D-1)\times D$ matrix is rank-deficient, then the intersection of the $D$ hypersurfaces (i.e., the solution of~(\ref{eq:VVc})) is a $p$-manifold (with $p\geq1$) on the $D$-dimensional space and does not generate a new cell.
Hence, combinations of $D$ rows of ${\bf V}$ that result in linearly dependent rows of the $(D-1)\times D$ matrix in~(\ref{eq:VVc}) can be simply ignored.}
Then, the index-set ${\mathcal I}^+$ that corresponds to a neighboring cell is computed by 
\begin{equation}
\text{top}_K({\bf V}{\bf c}\!\left({\boldsymbol\phi}\right)).
\label{eq:selectVc}
\end{equation}
Note that the $D$ intersecting hypersurfaces have the same value at $\boldsymbol\phi$.
Hence,~(\ref{eq:selectVc}) returns ambiguity regarding the sorting of these particular $D$ hypersurfaces.
If $d<D$ hypersurfaces of these belong to the $K$ largest ones, then, due to this ambiguity, we have to consider all $\binom{D}{d}$ combinations of $d$ hypersurfaces among the $D$ intersecting ones, where $\binom{D}{d}<\binom{D}{\left\lfloor\frac{D}{2}\right\rfloor}$.
Finally, we have to repeat the above procedure for all $2^{D-1}$ sign combinations in~(\ref{eq:VVc}) and any combination of $D$ intersecting hypersurfaces among the $N$ ones.
The total number of combinations is $\binom{N}{D}$, hence the cardinality of ${\mathcal S}$ is upper bounded by $2^{D-1}\binom{D}{\left\lfloor\frac{D}{2}\right\rfloor}\binom{N}{D}={\mathcal O}\left(N^D\right)$.

\begin{algorithm}[t!]
\caption{Computation of the sparse principal component of a rank-$D$ matrix with complexity ${\mathcal O}\left(N^{D+1}\right)$}
\begin{algorithmic}
\Require $\mathbf{V} \in{\mathbbm R}^{N \times D}$, $K \in [N]$
\State $ \mathcal{S} \gets \{\}$ {\it(set of candidate supports)}
\For{all $\binom{N}{D}$ sets $\lbrace i_1,\ldots, i_D \rbrace \subseteq [N]$}
\For{all $2^{D-1}$ sequences $(b_1,\ldots, b_{D-1})\in \lbrace \pm 1 \rbrace^{D-1}$}
    \State $ \mathbf{c} \leftarrow \text{nullspace}
\left(
\left[
\begin{array}{l@{}c@{}l}
{\bf V}_{i_1,:}&-&b_1{\bf V}_{i_2,:}\\
&\vdots&\\
{\bf V}_{i_1,:}&-&b_{D-1}{\bf V}_{i_D,:}\\
\end{array}
\right]
\right)$
\State $\mathcal{I} \gets \text{top}_K( \mathbf{V} \mathbf{c})$
\If {$\left|\mathcal{I}\right| = K$}
\State $\mathcal{S} \gets \mathcal{S} \cup \{{\mathcal I}\}$
\Else
\State $\mathcal{T} \gets \mathcal{I} - \lbrace i_1, \hdots, i_D \rbrace$
\State $r \gets  K - \left| \mathcal{T} \right|$
\For{all $\binom{D}{r}$ $r$-subsets $\mathcal{M} \subseteq \lbrace i_1,\ldots, i_D \rbrace$}
\State $\hat{\mathcal{I}} \gets \mathcal{T} \cup \mathcal{M}$
\State $\mathcal{S} \gets \mathcal{S} \cup \{\hat{\mathcal I}\}$
\EndFor
\EndIf
\EndFor
\EndFor
\State $\displaystyle\mathcal{I}_\text{opt} = \argmax_{\mathcal{I} \in \mathcal{S}}\sigma_{\max}(\mathbf{V}_{{\mathcal I},:})$
\State \Return $\mathcal{I}_\text{opt } \& \text{ principal left singular vector of }\mathbf{V}_{{\mathcal I}_\text{opt},:}$
\end{algorithmic}
\label{fig:algo_rankD}
\end{algorithm}

A pseudocode that includes all the above steps is presented in Algorithm~\ref{fig:algo_rankD}.
Its complexity to build ${\mathcal S}$ is determined by the complexity to build each element of it (i.e., each index-set ${\mathcal I}^+$) for each examined intersection through~(\ref{eq:selectVc}).
Note that function $\text{top}_K$ has complexity ${\mathcal O}(N)$ and the cardinality of ${\mathcal S}$ is ${\mathcal O}\left(N^D\right)$.
Hence, the overall complexity to build ${\mathcal S}$ is ${\mathcal O}\left(N^{D+1}\right)$.
Finally, we mention that the computation of each element of ${\mathcal S}$ is performed independently of each other.
Therefore, the proposed Algorithm~\ref{fig:algo_rankD} that builds ${\mathcal S}$ and solves~(\ref{initial_problem}) or, equivalently,~(\ref{eq:maxsigma}) with complexity ${\mathcal O}\left(N^{D+1}\right)$ is fully parallelizable and memory efficient.

\section{An Algorithm of Complexity ${\mathcal O}\left(N^2\log N\right)$ for Rank-$2$ Matrices}
\label{section:rank2}

In the special case of a rank-$2$ matrix ${\bf A}$ (i.e., $D=2$), Algorithm~\ref{fig:algo_rankD} computes the sparse principal component of ${\bf A}$ with complexity ${\mathcal O}\left(N^3\right)$ for any sparsity value.
In this section, we develop an algorithm that computes the sparse principal component of ${\bf A}$ with complexity ${\mathcal O}\left(N^2\log N\right)$.

\subsection{A serial algorithm for rank-$2$ matrices}
\label{subsection:rank2serial}

For the rank-$2$ case, Algorithm~\ref{fig:algo_rankD}, discussed in Subsection~\ref{subsection:rank2algo}, relied on identifying a polynomial number of non-overlapping intervals on $\Phi$, each associated with a candidate support set $\mathcal{I}$.
These intervals are induced by the pairwise intersections of curves in $|\mathbf{u}(\phi)|$.
The candidate support set associated with an interval is determined at the leftmost point of the interval.
The two-step algorithm first computes all $2{N \choose 2}$ pairwise intersection points.
In the second step, it determines $\mathcal{I}(\hat{\phi})$, the set of indices of the $K$ largest elements of $|\mathbf{u}(\hat{\phi})|$, at each intersection point $\hat{\phi}$ individually, in linear time.
In this section, we present an algorithmic enhancement that reduces the overall computational complexity, exploiting the correlation between candidate support sets of neighboring intervals.

The relative order of $\left| u_i(\phi)\right|$ and $\left| u_i(\phi)\right|$ changes only at the points where the $i$th and $j$th curves intersect.
Conversely, if those two are the only curves intersecting at a point $\hat{\phi}$, then the ordering of the remaining elements of $|\mathbf{u}(\phi)|$ in the intervals immediately preceding  and succeeding $\hat{\phi}$ is identical.
The limited change on ordering of the curves in $|\mathbf{u}(\phi)|$ across an intersection point $\hat{\phi}$ implies that the differences between the candidate support sets of the adjacent intervals cannot be arbitrary.
The support sets associated with two neighboring intervals differ in at most one element.
More formally, let $\hat{\phi}$ be the intersection point of the $i$th and $j$th curves of $|\mathbf{u}(\phi)|$.
There exists an $\epsilon>0$ such that $\hat{\phi}$ is the only intersection point lying in $[\hat{\phi}-\epsilon, \hat{\phi}+\epsilon]$.
If $|u_i(\phi)| < |u_j(\phi)|$ in $[\hat{\phi}-\epsilon, \hat{\phi})$, then $|u_i(\phi)| > |u_j(\phi)|$ in $(\hat{\phi}, \hat{\phi}+\epsilon]$, and vice versa.
The ordering of all other curves remains unaltered over $\hat{\phi}$.
If the $i$th and $j$th curves  are both members of the candidate support in the interval preceding $\hat{\phi}$, then $\mathcal{I}(\hat{\phi}+\epsilon)=\mathcal{I}(\hat{\phi}-\epsilon)$.
The same holds if neither  is included in $\mathcal{I}(\hat{\phi}-\epsilon)$.
On the contrary, if exactly one of $i$ and $j$ belongs to $\mathcal{I}(\hat{\phi}-\epsilon)$, then the candidate sets associated with the two neighboring intervals differ in exactly one element.
If, w.l.o.g., the $i$th curve is the one belonging to $\mathcal{I}(\hat{\phi}-\epsilon)$, then $\mathcal{I}(\hat{\phi}+\epsilon) = \left(\mathcal{I}(\hat{\phi}-\epsilon)-\{i\} \right) \cup \{j\}$.

The key observation is that, if the candidate support set associated with a particular interval is known, then the set associated with a neighboring interval can be determined with a constant complexity.
The above observations readily suggest the following procedure for determining the candidate support sets:
\begin{enumerate}
\item
\label{serial:step-find}
Compute all $2{N \choose 2}$ pairwise intersection points of the curves in $\left| \mathbf{u}(\phi)\right|$.
\item
\label{serial:step-sort}
Sort the intersection points in increasing order.
Let $\phi_1, \phi_2, \ldots,\phi_{2{N \choose 2}}$ be the sorted sequence.
\item
\label{serial:step-init}
Determine $\mathcal{I}(\phi_1)$, the set of indices of the $K$ largest curves in $|\mathbf{u}(\phi)|$ at the first intersection point $\phi_1$.
\item
\label{serial:step-scan}
Successively determine $\mathcal{I}(\phi_t)$ at consecutive intersection points.
At $\phi_t$, the candidate support set $\mathcal{I}(\phi_t)$ is determined by appropriately updating  $\mathcal{I}(\phi_{t-1})$, the set associated with the previous interval.
\end{enumerate}
Once all candidate support sets have been collected, the optimal solution is determined as in Algorithm~\ref{fig:algo_rankD}.

\begin{figure}[t!]
\centerline{\epsfig{file=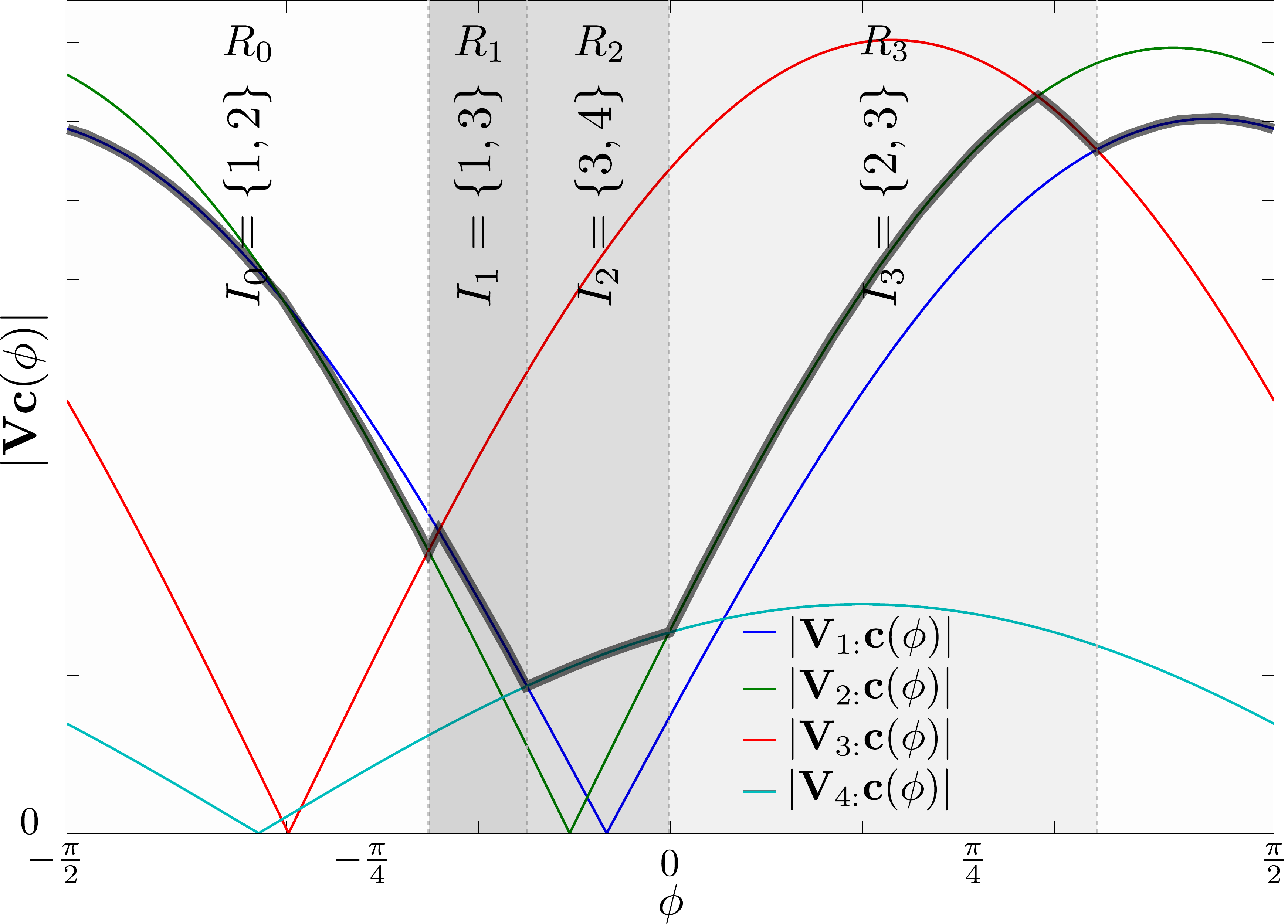,width=\myfigurewidth}}
\caption{Execution of Algorithm~\ref{algo:rank2-serial-algo-simple} on an arbitrary $4\times2$ matrix ${\bf V}$ for sparsity $K=2$.}
\label{fig:rank2-serial-V4x2-example}
\end{figure}

An illustrative figure that explains the above steps is presented in Fig.~\ref{fig:rank2-serial-V4x2-example}, using the same $4\times2$ matrix ${\bf V}$ as the one examined in Section~\ref{section:parallel} (Figs.~\ref{fig:normalcells} and~\ref{fig:rank2-serial-V4x2-example}).
We seek the optimal $2$-sparse solution, i.e., $K=2$.
The algorithm starts at $\phi=-\pi/2$ and scans $\Phi$, keeping track of the $K$th order curve, highlighted with a black solid line.
Vertical dashed lines indicate the intersections points at which the support of the optimal $\mathbf{x}$ changes.
Regions of $\Phi$ corresponding to different support have a different background color.
The optimal support in each region is depicted in the top of the figure.
Note that 2 of the ${N \choose 2} = 6$ support sets need not be considered: sets $\lbrace{ 1,4\rbrace}$ and $\lbrace 2, 4 \rbrace$ do not appear for any $\phi\in\Phi$.

\begin{figure}[t!]
\centerline{\epsfig{file=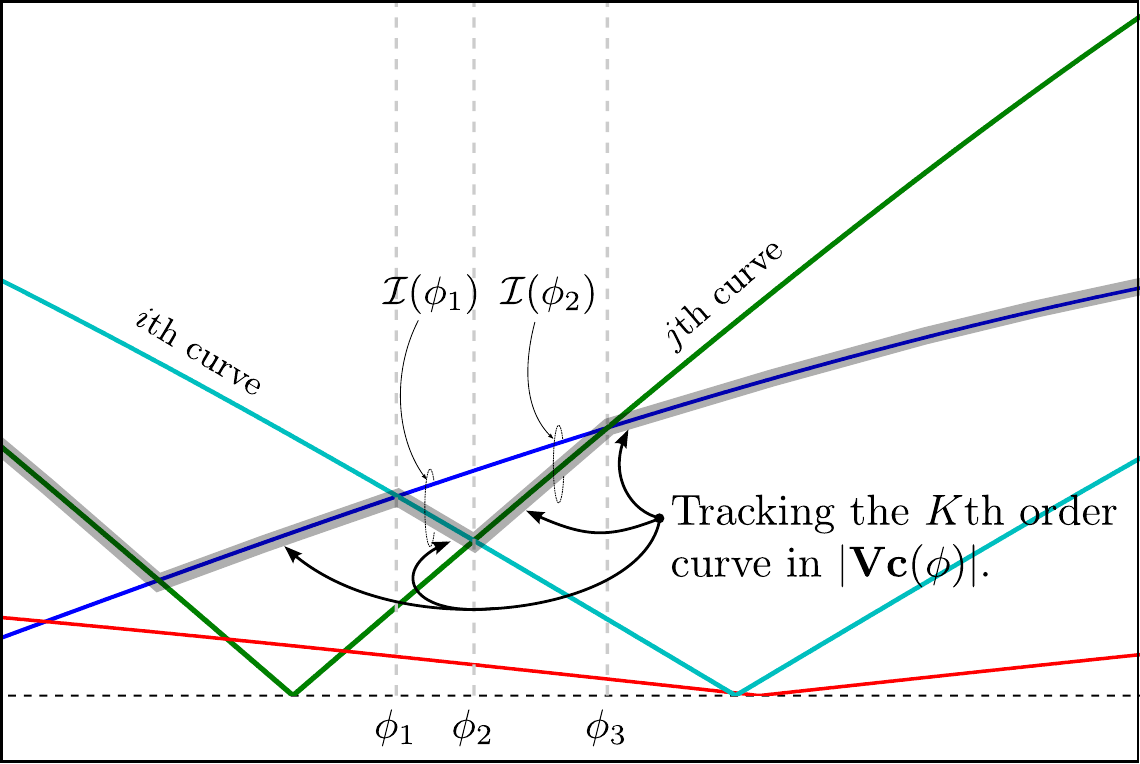,width=\myfigurewidth}}
\caption{Curves in $\left|\mathbf{u}(\phi) \right|$ in a constrained region of $\Phi$ for an arbitrary $4\times2$ matrix ${\bf V}$.
We highlight the evolution of the $K$th order curve, for $K=2$.
Angles $\phi_1$, $\phi_2$, and $\phi_3$ are three consecutive intersection points.
The candidate support sets in $[\phi_1, \phi_2)$ and $[\phi_2, \phi_3)$ differ by exactly one element.
The $K$th order curve changes over the intersection point $\phi_2$.}
\label{fig:rank_2_explain_modified_serial_algo}
\end{figure}

A pseudocode that implements the above serial algorithmic steps is presented in Algorithm~\ref{algo:rank2-serial-algo-simple}.
It improves upon the computational complexity of its parallel counterpart (Algorithm~\ref{fig:algo_rankD}) circumventing the $O(N)$ construction of the candidate set at each individual intersection point.
The computation and sorting of all intersection points (steps \ref{serial:step-find} and \ref{serial:step-sort}) is performed in $O(N^2\log N)$ operations.
The construction of $\mathcal{I}(\phi_1)$ is performed in linear time.
Finally, each successive update in the last step requires $O(1)$ operations.
Therefore, the overall complexity of the serial Algorithm~\ref{algo:rank2-serial-algo-simple} is $O\left( N^2\log N\right)$ as opposed to the complexity $O\left(N^3\right)$ of the parallel Algorithm~\ref{fig:algo_rankD} for $D=2$.
The disadvantage of Algorithm~\ref{algo:rank2-serial-algo-simple} is that it is not parallelizable.

\begin{algorithm}[t!]
\caption{Serial computation of the sparse principal component of a rank-$2$ matrix with complexity ${\mathcal O}\left(N^2\log N\right)$}
\begin{algorithmic}
\Require $\mathbf{V} \in{\mathbbm R}^{N \times 2}$, $K$
\State $ \mathcal{S} \gets \{\}$ {\it(set of candidate supports)}
\State $\boldsymbol{\phi} \gets \text{sort} \left( \text{all } 2{N \choose 2}\text{ intersection points} \right)$\\
{\it(remember the pair $\lbrace i,j\rbrace$ of curves intersecting at each $\phi_t$)}
\State $\mathcal{I}_0 \gets \text{top}_K \left( \mathbf{V}\mathbf{c}(\phi_0) \right) $ at $\phi_0 = -\frac{\pi}{2}$
\For{$t = 1:\text{length}(\boldsymbol{\phi})$}
\State $\lbrace i,j \rbrace \gets$ curves intersecting at $\phi_t$
\State Compute $\mathcal{I}_t$  by modifying $\mathcal{I}_{t-1}$ in $\mathcal{O}(1)$:
\If{both or none of $i$ and $j$ are in $\mathcal{I}_{t-1}$,}
\State$\mathcal{I}_t \gets \mathcal{I}_{t-1}$
\ElsIf{only $i$ (or $j$) is in $\mathcal{I}_{t-1}$,}
\State $\mathcal{I}_t \gets \left(\mathcal{I}_{t-1} - \lbrace i \; (\text{or } j)\rbrace\right) \cup \lbrace j \; (\text{or } i)\rbrace$
\EndIf
\State $\mathcal{S} \gets \mathcal{S} \cup \{{\mathcal I}_t\}$
\EndFor
\State $\displaystyle\mathcal{I}_\text{opt} = \argmax_{\mathcal{I} \in \mathcal{S}}\sigma_{\max}(\mathbf{V}_{{\mathcal I},:})$
\State \Return $\mathcal{I}_\text{opt } \& \text{ principal left singular vector of }\mathbf{V}_{{\mathcal I}_\text{opt},:}$
\end{algorithmic}
\label{algo:rank2-serial-algo-simple}
\end{algorithm}

\subsection{A modified serial algorithm for rank-$2$ matrices}

Algorithms~\ref{fig:algo_rankD} and~\ref{algo:rank2-serial-algo-simple} described so far collect the candidate support sets by examining all pairwise intersection points.
The number of distinct candidate support sets, however, can be significantly less: multiple intervals on $\Phi$ may be associated with the same support set.
In the following, we describe a simple modification of Algorithm~\ref{algo:rank2-serial-algo-simple} that aims at reducing the total number of intersection points computed and examined.
Although the worst case complexity remains the same, the modified serial algorithm may significantly speed up execution in certain cases.

Consider three consecutive intersection points $\phi_1$, $\phi_2$, and $\phi_3$, such that the candidate support sets on the two sides of $\phi_2$ are different (see, for example, Fig.~\ref{fig:rank_2_explain_modified_serial_algo}).
Let $\mathcal{I}(\phi_1)$ denote the set of indices of the $K$ largest curves in $\left|\mathbf{u}(\phi)\right|$ in the interval $[\phi_1, \phi_2)$.
Similarly, $\mathcal{I}(\phi_2)$ is the set associated with $[\phi_2, \phi_3)$.
The two sets differ by one element:
if $\mathcal{C} = \mathcal{I}(\phi_1) \cap \mathcal{I}(\phi_2)$ is the set of the $K-1$ common elements, then $\mathcal{I}(\phi_1) = \mathcal{C} \cup \{i\}$ and $\mathcal{I}(\phi_2) = \mathcal{C} \cup \{j\}$, for some curves $i$ and $j$.

At $\phi_2$, over which the candidate support set changes, the two curves intersect.
The $j$th curve joins the set of $K$ largest curves of $\left|\mathbf{u}(\phi)\right|$, displacing the $i$th curve which was a member of $\mathcal{I}(\phi_1)$.
In particular, the $i$th curve must be the smallest element of $\left|\mathbf{u}(\phi)\right|$ in $[\phi_1, \phi_2)$.
To see that, assume for the sake of contradiction that among the curves in $\mathcal{I}(\phi_1)$, the $l$th curve was the smallest one in $[\phi_1, \phi_2)$, where $l \neq i$.
Then, $u_i(\phi) > u_l(\phi) > u_j(\phi)$ in $[\phi_1, \phi_2)$.
By assumption, $u_i(\phi) = u_j(\phi)$ at $\phi_2$.
Due to the continuity of the curves, there must exist a point in  $[\phi_1, \phi_2)$ where either $u_i(\phi) = u_l(\phi)$ or $u_l(\phi) = u_j(\phi)$.
However, no intersection point lies in $[\phi_1, \phi_2)$.
Following a similar argument, the $j$th curve must be the largest curve in $[\phi_1, \phi_2)$ among those not included in $\mathcal{I}(\phi_1)$.
Moreover, the $j$th curve becomes the $K$th-order element of $\left|\mathbf{u}(\phi)\right|$ in $[\phi_2, \phi_3)$, i.e., it is the smallest curve in $\mathcal{I}(\phi_2)$.

The key observation is that, along $\Phi$, the candidate support set $\mathcal{I}(\phi)$ changes only at the intersection points of the $K$th-order curve in $\left| \mathbf{u}(\phi)\right|$.
Assume that the candidate support set $\mathcal{I}(\phi_0)$ is known at $\phi_0 = -\frac{\pi}{2}$ and the $i$th curve is the $K$th-order curve in $\left| \mathbf{u}(\phi)\right|$, i.e., the smallest curve in $\mathcal{I}(\phi_0)$.
Moving along $\Phi$, the first point where the candidate support set can potentially change is the closest intersection point of the $i$th curve.
Let $\phi_1 \in [\phi_0, \frac{\pi}{2})$ be that point and $j$ be the intersecting curve.
If $j \notin {\mathcal{I}(\phi_0)}$, then the $j$th curve joins the set $\mathcal{I}(\phi_1)$ at $\phi_1$, displacing the $i$th curve.
If $j \in {\mathcal{I}(\phi_0)}$, then $\phi_1$ is only a point of internal reordering for $\mathcal{I}(\phi_0)$, hence $\mathcal{I}(\phi_1) = \mathcal{I}(\phi_0)$.
In either case, however, the $j$th curve becomes the $K$th-order curve immediately after $\phi_1$.
Proceeding in a similar fashion, the next point where the candidate support set can potentially change is a point $\phi_2 \in [\phi_1, \frac{\pi}{2})$ closest to $\phi_1$, where the $j$th curve intersects one of the other $N$ curves.

\begin{algorithm}[t!]
\caption{Modified serial computation of the sparse principal component of a rank-$2$ matrix with complexity ${\mathcal O}\left(N^2\log N\right)$}
\begin{algorithmic}
\Require $\mathbf{V} \in{\mathbbm R}^{N \times 2}$, $K$.
\State $\phi_0 \gets -\frac{\pi}{2}$,  $t \gets 0$
\State $ \mathcal{S} \gets \{\}$ {\it(set of candidate supports)}
\State $\mathcal{I}_0 \gets \text{top}_{K}\left(\mathbf{V}\mathbf{c}(\phi_0)\right)$,
\State $i\gets$ index of the smallest magnitude curve in $\mathcal{I}_0$.
\Loop
\State $t \gets t+1$.
\State $\mathcal{H} \gets $ points of intersection of curve $i$ with all other
\State \phantom{$\mathcal{H} \gets$} curves in $\left(\phi_{t-1}, \frac{\pi}{2}\right)$, if not already computed
\If {$\mathcal{H}= \emptyset$,}
\State exit loop
\EndIf
\State  $\phi_t \gets\min\{{\mathcal H}\}$
\State $j \gets$ index of the curve intersecting curve $i$ at $\phi_t$
\If{$j \in \mathcal{I}_{t-1}$,}
\State $\mathcal{I}_{t} \gets \mathcal{I}_{t-1}$
\Else
\State  $\mathcal{I}_{t} \gets \left( \mathcal{I}_{t-1} - \{i\}\right) \cup \lbrace j \rbrace$
\EndIf
\State $\mathcal{S} \gets \mathcal{S} \cup \{{\mathcal I}_{t}\}$
\State $i\gets  j $ {\it(the new $K$th order curve)}
\EndLoop
\State $\displaystyle\mathcal{I}_\text{opt} = \argmax_{\mathcal{I} \in \mathcal{S}}\sigma_{\max}(\mathbf{V}_{{\mathcal I},:})$
\State \Return $\mathcal{I}_\text{opt } \& \text{ principal left singular vector of }\mathbf{V}_{{\mathcal I}_\text{opt},:}$
\end{algorithmic}
\label{algo:rank2-serial-algo-modified}
\end{algorithm}

A pseudocode that implements the above steps and modifications is presented in Algorithm~\ref{algo:rank2-serial-algo-modified}.
In summary, instead of computing all pairwise intersections at the first step, Algorithm~\ref{algo:rank2-serial-algo-modified} postpones the computation of the intersection points of the $i$th curve until the latter becomes the $K$th-order curve in $\left| \mathbf{u} (\phi) \right|$.
The motivation behind this enhancement lies on the fact that multiple curves might never become $K$th in order.

\begin{figure}[t!]
\centerline{\epsfig{file=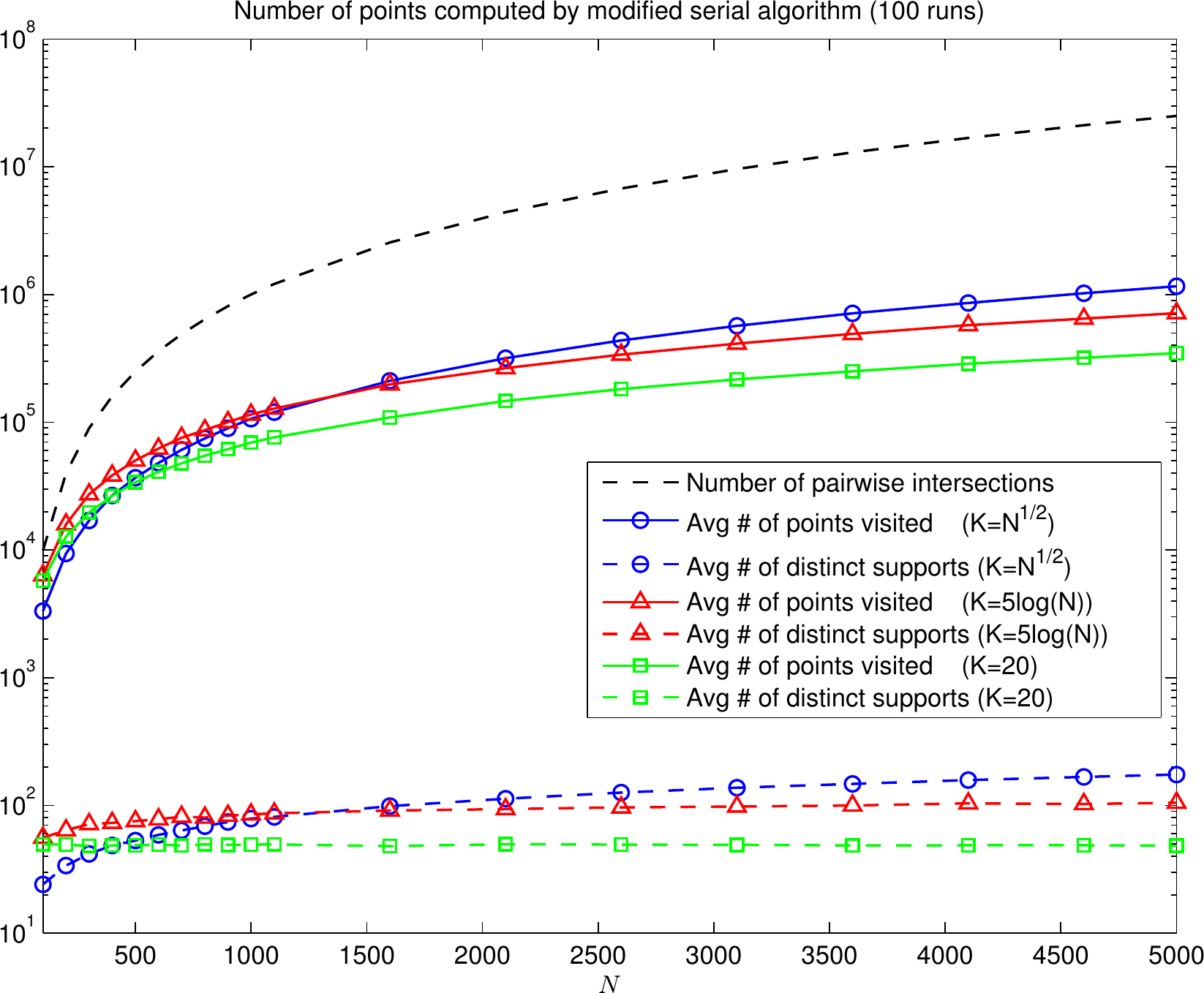,width=\myfigurewidth}}
\caption{Average number of intersection points and distinct candidate supports computed by Algorithm~\ref{algo:rank2-serial-algo-modified} for sparsity $K=\sqrt{N}$, $5\log N$, and $20$ and total number of intersection points $2\binom{N}{2}$, as a function of $N$.}
\label{fig:rank_2_complexity_fig_02}
\end{figure}

To justify this, in Fig.~\ref{fig:rank_2_complexity_fig_02}, we present the average number of intersection points computed%
\footnote{Although even fewer points will be eventually visited, the complexity of Algorithm~\ref{algo:rank2-serial-algo-modified} is dominated by the number of intersections computed.}
by Algorithm~\ref{algo:rank2-serial-algo-modified} as a function of $N$, for sparsity $K=\sqrt{N}$, $5\log N$, and $20$.
The average is estimated over $100$ independent instances of the $N\times2$ matrix ${\mathbf V}$.
We also present the average number of distinct candidate supports for the three different values of $K$ and the total number of intersection points $2 \binom{N}{2}$.
We observe that Algorithm~\ref{algo:rank2-serial-algo-modified} computes noticeably fewer points than the total number of $2\binom{N}{2}$ pairwise intersection points, indicating that a significant fraction of the curves never rises to the $K$th order.
Finally, we note that the average number of distinct candidate supports, i.e., the cardinality of ${\mathcal S}$, is significantly smaller than the number of intersection points computed by Algorithm~\ref{algo:rank2-serial-algo-modified}, revealing potential for further reduction of the computational complexity of the algorithms developed in this present work.

\section{Conclusions}
\label{section:conclusions}

We proved that the sparse principal component of an $N\times N$ matrix is computable with complexity ${\mathcal O}\left(N^{D+1}\right)$ if the matrix is positive semidefinite and its rank $D$ is constant.
This holds true for any sparsity value $K$ (that is, even if $K$ grows with $N$).
Our constructive proof was accompanied by a fully parallelizable and memory efficient algorithm which computes the sparse principal component with complexity ${\mathcal O}\left(N^{D+1}\right)$.
For the special case of rank-$2$ matrices, we developed alternative serial algorithms of complexity ${\mathcal O}\left(N^2\log N\right)$.
The construction steps and properties of the algorithms that we presented in this work indicate that implementations of even lower complexity may be possible.

\begin{IEEEbiographynophoto}{Megasthenis Asteris} (S'12) received the Diploma degree (five-year program) in electronic and computer engineering from the Technical University of Crete, Chania, Greece, in 2010 and the M.Sc. degree in electrical engineering from the University of Southern California, Los Angeles, in 2012.
He is currently working toward the Ph.D. degree at the Department of Electrical and Computer Engineering, University of Texas at Austin.
His research interests are in coding theory with an emphasis on distributed storage and large-scale data processing.
\end{IEEEbiographynophoto}

\begin{IEEEbiographynophoto}{Dimitris S. Papailiopoulos} (S'10) received the Diploma (five-year program) and M.Sc. degrees in electronic and computer engineering from the Technical University of Crete, Greece, in 2007 and 2009, respectively.
Between 2009 and 2012, he was a graduate student with the Department of Electrical Engineering, University of Southern California, Los Angeles.
Since January 2013, he has been with the Department of Electrical and Computer Engineering, University of Texas at Austin, where he is currently working towards the Ph.D. degree.
His research interests are in the areas of coding theory, information theory, and large-scale data processing with an emphasis on codes for distributed storage, large-scale optimization for data and graph analysis, and succinct data representations.
\end{IEEEbiographynophoto}

\begin{IEEEbiographynophoto}{George N. Karystinos} (S'98-M'03) was born in Athens, Greece, on April 12, 1974.
He received the Diploma degree in computer science and engineering (five-year program) from the University of Patras, Patras, Greece, in 1997 and the Ph.D. degree in electrical engineering from the State University of New York at Buffalo, Amherst, NY, in 2003.

From 2003 to 2005, he held an Assistant Professor position in the Department of Electrical Engineering, Wright State University, Dayton, OH.
Since 2005, he has been with the Department of Electronic and Computer Engineering, Technical University of Crete, Chania, Greece, where he is presently an Associate Professor.
His research interests are in the general areas of $L_1$-norm principal component analysis of data and signals, communication theory, and adaptive signal processing with applications to signal waveform design, low-complexity sequence detection, and secure wireless communications.

Dr. Karystinos is a member of the IEEE Signal Processing, Communications, Information Theory, and Computational Intelligence Societies.
For articles that he coauthored with students and colleagues, he received a 2001 IEEE International Conference on Telecommunications best paper award, the 2003 IEEE \textsc{Transactions On Neural Networks} Outstanding Paper Award, the 2011 IEEE International Conference on RFID-Technologies and Applications Second Best Student Paper Award, and the 2013 International Symposium on Wireless Communication Systems Best Paper Award in Signal Processing and Physical Layer Communications.
\end{IEEEbiographynophoto}

\vfill

\end{document}